\documentclass[twocolumn,aps,prl,showpacs]{revtex4}

\usepackage{epsfig}
\usepackage{graphicx}
\usepackage{amsmath}
\usepackage{bm}

\begin{document}

\title{Novel Josephson Effect in
Triplet Superconductor - Ferromagnet - Triplet Superconductor Junctions}
\author{Boris Kastening$^1$, Dirk K.~Morr$^{1,2}$, Dirk Manske$^3$ and
Karl Bennemann$^1$} \affiliation{$^1$ Institut f\"ur Theoretische
Physik,
Freie Universit\"{a}t Berlin, 14195 Berlin, Germany \\
$^2$ Department of Physics, University of Illinois at Chicago,
Chicago, IL \\ $^3$ Max-Planck-Institut f\"ur Festk\"orperforschung,
Heisenbergstrasse 1, 70569 Stuttgart, Germany}

\date{\today}
\begin{abstract}
We predict a novel type of Josephson effect to occur in triplet
superconductor - ferromagnet - triplet superconductor Josephson
junctions. We show that the Josephson current, $I_J$, exhibits a
rich dependence on the relative orientation between the
ferromagnetic moment and the ${\bf d}$-vectors of the
superconductors. This dependence can be used to build several types
of {\it Josephson current switches}. Moreover, we predict an
unconventional temperature dependence of $I_J$ in which $I_J$
changes sign with increasing temperature.

\end{abstract}


\maketitle

Josephson junctions made of unconventional superconductors have
attracted significant interest over the last few years due to their
unconventional quantum transport properties
\cite{reviews,anomaly,Rie98,Asa03,Vac03,Kwon04}. The latter are
determined by the formation of low-energy Andreev bound states
\cite{Hu94}, which, for example, lead to a low-temperature anomaly
in the Josephson current \cite{anomaly,Rie98}. Josephson junctions
consisting of triplet superconductors are of particular interest
since the superconductors' odd spatial symmetry guarantees the
formation of low-energy Andreev states for any orientation of the
superconductors in the junction \cite{Asa03,Vac03,Kwon04}.

%
%
\begin{figure}[h]
\includegraphics[width=7.0cm,angle=0]{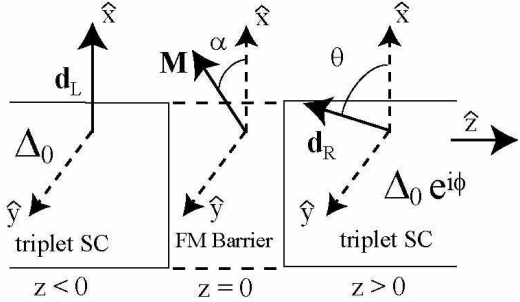}
\caption{Schematic picture of a 1D Triplet Superconductor -
Ferromagnet - Triplet Superconductor (TSFTS) Josephson Junction.}
\label{Junction}
\end{figure}
In this Letter, we predict a novel type of Josephson effect in
one-dimensional (1D) triplet superconductor - ferromagnet - triplet
superconductor (TSFTS) junctions (a schematic picture of such a
junction is shown in Fig.~\ref{Junction}). In particular, we
demonstrate a rich dependence of the Josephson current, $I_J$, on
the relative orientation between the ferromagnetic moment, ${\bf
M}$, in the barrier and the ${\bf d}_{L,R}$-vectors of the left and
right triplet superconductor \cite{comment1}, as described by the
angles $\alpha$ and $\theta$ (see Fig.~\ref{Junction}). This
dependence can be used to create {\it Josephson current switches} in
which small changes of $\alpha$ or $\theta$ can tune the junction
between two ``current states", in which $I_J$ is either ``on" ($I_J
\not = 0$) or ``off" ($I_J \approx 0$), or differs in its direction.
These types of two-level systems are of great current interest in
the field of quantum information technology \cite{Kane98}. Moreover,
we predict that a TSFTS junction leads to a {\it qualitatively} new
temperature dependence of $I_J$ such that in certain cases, $I_J$
changes sign (i.e., its direction) with increasing temperature.
Finally, we show that adiabatic changes in the orientation of ${\bf
d}_{L,R}$ yield a behavior of $I_J$ which is significantly altered
from its equilibrium form.

We take the 1D TSFTS junction to be aligned along the $z$-axis and
be described by the Hamiltonian $H=\int dz \, dz^\prime
H(z,z^\prime)$, where ($\hbar=1$)
\begin{eqnarray}
\lefteqn{H(z,z^\prime) = \sum_\sigma \psi^\dagger_\sigma (z^\prime)
\delta(z-z^\prime) \left[
-\frac{\partial_z^2}{2m}-\mu \right] \psi_\sigma (z)}
\nonumber \\
&&{}
+ \Big\{\frac{\Delta(z,z^\prime)}{2} \left[e^{-i \theta_j}
\psi^\dagger_\uparrow (z^\prime) \psi^\dagger_\uparrow (z) -e^{i
\theta_j} \psi^\dagger_\downarrow (z^\prime)
\psi^\dagger_\downarrow(z)
\right]
\nonumber \\
&&\qquad{} +\text{h.c.} \Big\} - {\bf M}(z,z^\prime)\cdot
\sum_{\alpha,\beta} \psi^\dagger_\alpha (z^\prime) {\hat
\sigma}_{\alpha \beta} \psi_\beta (z), \label{H1}
\end{eqnarray}
and $\psi^\dagger_\sigma (z)$ and $ \psi_\sigma (z)$ are the
fermionic creation and annihilation operators for a particle with
spin $\sigma$ at site $z$, respectively, ${\hat \sigma}$ are the
Pauli matrices, and $\Delta(z,z^\prime)=-\Delta(z^\prime,z)$ is the
superconducting gap. The ${\bf d}$-vector of the triplet
superconductors on the left ($j=L, z<0$) and right ($j=R, z>0$) of
the barrier is given by ${\bf d}_{j}=(\cos \theta_j, \sin
\theta_j,0)$. We take $\theta_L \equiv 0$ such that ${\bf d}_L
\parallel {\hat {\bf x}} $ and $\theta_R = \theta$ with ${\bf
d}_R$ lying in the spin $xy$-plane. The ferromagnetic junction,
located at $z=0$, possesses a moment $M_0$ and represents a magnetic
scattering potential, described by the last term in Eq.(\ref{H1}),
with ${\bf M}(z,z^\prime)=M_0 (\cos \alpha, \sin \alpha,0)
\delta(z)\delta(z^\prime)$.

We show below that two Andreev bound states with energies $E_{a,b}$
are formed in the TSFTS junction. The Josephson current flows
through these two states \cite{com2} and is given by \cite{Zag98}
\begin{equation}
I_J=I_J^a+I_J^b=-\frac{e}{\hbar} \sum_{i=a,b} \frac{\partial
E_i}{\partial \Phi} \tanh \left( \frac{E_i}{2 k_B T} \right).
\end{equation}
In order to obtain the energies of the Andreev states, we start from
Eq.(\ref{H1}) and derive the Bogoliubov-de Gennes (BdG) equations
\cite{Zag98,Gen89} by introducing the unitary Bogoliubov
transformation
\begin{eqnarray*}
\psi_\uparrow (z)&\!=\!&\sum_n u_n(z) a_n  + v_n(z)
b_n^\dagger+w_n(z) b_n + x_n(z) a_n^\dagger, \\
\psi_\downarrow (z)&\!=\!&\sum_n -u_n(z) b_n + v_n(z)
a_n^\dagger-w_n(z) a_n + x_n(z) b_n^\dagger,
\end{eqnarray*}
where the sum runs over all eigenstates of the junction. For the
wave functions of the localized Andreev states on the left and right
of the barrier, we make the ansatz
\begin{equation}
\Psi_j(z)=
\begin{pmatrix}
u_j(z)\\
v^{\ast}_j(z)\\
w_j(z)\\
x^{\ast}_j(z)
\end{pmatrix}
=e^{c_j \kappa z} \sum_{\gamma=\pm} A_{j,\gamma}
\begin{pmatrix}
u_{j,\gamma}\\
v^{\ast }_{j,\gamma}\\
w_{j,\gamma}\\
x_{j,\gamma}^{\ast }
\end{pmatrix}
e^{\gamma i k_F z},
\end{equation}
where $c_j=+1 (-1)$, $j=L (R)$, and $\kappa^{-1}$ is the decay
length of the Andreev state. Using for definiteness a $p_z$-wave
symmetry of the superconducting gap, $\Delta(k_z)=\Delta_j \sin k_z$
(with lattice constant set to unity), the BdG equations take the
form $\hat{H} \Psi_j(z) = E \Psi_j(z)$ where
$$
\hat{H}=
\begin{pmatrix}
-\frac{\partial_z^{2}}{2m}-\mu & ie^{-i\theta }\Delta _{j}\partial_z &
e^{-i\alpha }M_0\delta(z)\,\, & 0 \\
ie^{i\theta }\Delta _{j}^{\ast }\partial_z & \frac{\partial_z^{2}}{2m}
+\mu & 0 & e^{i\alpha }M_0\delta(z) \\
e^{i\alpha }M_0\delta(z)\, & 0 & -\frac{\partial_z^{2}}{2m}-\mu \, &
ie^{i\theta
}\Delta _{j}\partial_z \\
0 & e^{-i\alpha }M_0\delta(z) & ie^{-i\theta }\Delta _{j}^{\ast }\partial_z
&
\frac{\partial_z^{2}}{2m}+\mu
\end{pmatrix}.
$$
In the following we take $\Delta_L=\Delta_0$ and $\Delta_R=\Delta_0
e^{i \phi}$ for the superconducting gap on the left and right side
of the junction. The above eigenvalue equation for $E$ is subject to
the boundary conditions $\Psi_L(0)=\Psi_R(0)$ and
\begin{equation}
\partial_z \Psi_R(0) - \partial_z \Psi_L(0) = -2 m M_0
\begin{pmatrix}
0 & {\hat P}  \\
{\hat P}^\dagger & 0 \\
\end{pmatrix}
\Psi_R(0),
\end{equation}
where ${\hat P}= {\hat \sigma}_3 \cos \alpha \  -i \hat{ \sigma}_0
\sin \alpha$. The solution of the BdG equations yields two Andreev
states with energies
\begin{equation}
\frac{E_{a,b}}{ \Delta_0}=k_F \sqrt{D} \sqrt{ A+B-2C \pm 2
\sqrt{(A-C)(B-C)}}, \label{Eb}
\end{equation}
where
\begin{eqnarray*}
A&=& \cos^2(\phi/2) \left[1-D\sin^2(\theta/2) \right],  \\
B&=&  \sin^2(\theta/2) \left[1-D \cos^2(\phi/2)\right], \\
C&=& (1-D) \left[\cos^2(\phi/2) - \cos^2(\theta/2) \right]
\cos^2(\alpha-\theta/2),
\end{eqnarray*}
and $D=[1+g^2]^{-1}$ with $g=m M_0/k_F$. The two Andreev states
appear in the local density of states near the junction as two
particle-like and two hole-like peaks at energies $\mp E_{a,b}$,
respectively.

\begin{figure}[t]
\includegraphics[width=8.5cm,angle=0]{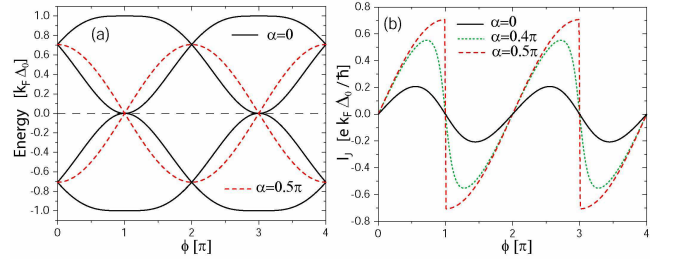}
\caption{(a) $\pm E_{a,b}$ and (b)$I_J$ as a function of $\phi$ for
$T=0$, $\theta=0$, $D=0.5$ and several values of $\alpha$.}
\label{parallel_d_phi}
\end{figure}

We first consider ${\bf d}_L \parallel {\bf d}_R$, and present in
Fig.~\ref{parallel_d_phi} the energies of the Andreev states and the
resulting Josephson current at $T=0$ as a function of $\phi$ for
several values of $\alpha$. Note that for $T=0$, only the negative
energy branches of the bound states are populated, and thus
contribute to $I_J$.
%
%
The dependence of $E_{a, b}$ on $\phi$ is {\it qualitatively}
different for ${\bf M} \parallel {\bf d}_{L,R}$ and ${\bf M} \perp
{\bf d}_{L,R}$. For ${\bf M} \perp {\bf d}_{L,R}$ ($\alpha=\pi/2$)
the Andreev states are degenerate with $E_{a,b}/\Delta_0=
k_F\sqrt{D}\cos{\phi/2}$, and possess well defined spin quantum
numbers $\sigma_a=\uparrow$, $\sigma_b=\downarrow$, since they are
not coupled by the scattering at the ferromagnetic barrier. The zero
energy level crossing at $\phi_{LC}=(2n+1) \pi$ ($n$=integer) occurs
with $\partial E_{a,b}/\partial \phi \not = 0$, resulting in
discontinuous jumps of $I_J$, as shown in
Fig.~\ref{parallel_d_phi}(b). It is interesting to note that
$E_{a,b}$ is identical to that of Andreev states near a potential
scattering barrier \cite{Kwon04}. In contrast, for $\alpha \not
=\pi/2$ the ferromagnetic barrier couples the Andreev states,
yielding a splitting of their energies, as shown in
Fig.~\ref{parallel_d_phi}(a) for ${\bf M}
\parallel {\bf d}_{L,R}$ ($\alpha=0$). This coupling yields
$\partial E_{a,b}/\partial \phi = 0$ at $\phi_{LC}$, such that $I_J$
evolves continuously with $\phi$ [Fig.~\ref{parallel_d_phi}(b)].
While these results remain qualitatively unchanged with decreasing
$D$, the dependence of $I_J$ on $D$ changes with the orientation of
${\bf M}$ and ${\bf d}_{L,R}$. Specifically, for $D \ll 1$ one
finds, to leading order in $D$, $I_J \sim D^{1/2}$ for $\alpha \not
= 0$, but $I_J \sim D^{3/2}$ for $\alpha=0$. Finally, note that for
$\phi=0$ and any $\alpha$, the induced Andreev states with $E_{a,
b}/ \Delta_0=k_F\sqrt{D}$ are identical to the impurity (Shiba)
states \cite{Shiba68} induced by a single (static) magnetic impurity
in a 1D triplet superconductor.

\begin{figure}[h]
\includegraphics[width=8.5cm,angle=0]{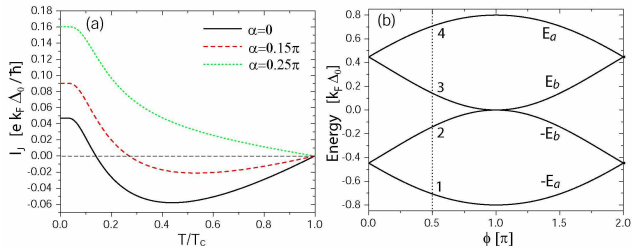}
\caption{(a) $I_J$ as a function
of $T/T_c$ for $\theta=0$, $\phi=0.5\pi$ and $D=0.2$. (b) $\pm
E_{a,b}$ as a function of $\phi$ for $\alpha=0$. The dotted line
corresponds to the parameters in (a).} \label{IJT}
\end{figure}

The Josephson current in a TSFTS junction exhibits an unconventional
temperature dependence in that $I_J$ can change sign, and thus its
direction, with increasing temperature, as shown in
Fig.~\ref{IJT}(a) for $\theta=0$ and $\phi=\pi/2$ (we assumed a BCS
temperature dependence of the superconducting gap).
%
%
In order to understand this sign change, we consider the
$\phi$-dependence of $E_{a,b}$ shown in Fig.~\ref{IJT}(b) and note
that at $T=0$, only the branches indicated by $1$ and $2$, belonging
to Andreev states $a$ and $b$, respectively, are occupied. Since the
derivatives $\partial E_{a,b}/\partial \phi$ possess opposite signs
for the two Andreev states, the corresponding currents through them,
$I_J^{a}<0$ and $I_J^{b}>0$, flow in opposite directions with
$I_J^{b} > |I_J^{a}|$. With increasing temperature, the occupation
of branches $2$ and $3$ changes more rapidly than those of branches
$1$ and $4$.  As a result, the magnitude of $I_J^b$ decreases more
quickly than that of $I_J^a$, and the total current, $I_J$,
eventually changes sign. Moreover, the qualitative nature of the
temperature dependence can be altered via a rotation of ${\bf M}$.
Specifically, as $\alpha$ increases, the $T=0$ value of $I_J^a$
decreases while $I_J^{b}$ remains practically unchanged. This leads
to a qualitative change in the temperature dependence of $I_J$ such
that for $\alpha \geq 0.21 \pi$, $I_J$ does not undergo a sign
change with increasing temperature [Fig.~\ref{IJT}(a)] .

\begin{figure}[h]
\includegraphics[width=8.5cm,angle=0]{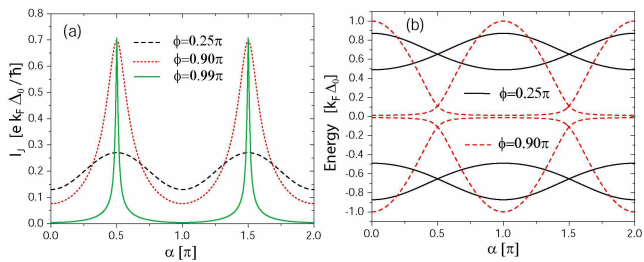}
\caption{(a) $I_J$ and (b) $\pm E_{a,b}$, as a function of $\alpha$
for $D=0.5$ at $T=0$, $\theta=0$ and several values of $\phi$.}
\label{parallel_d_alpha}
\end{figure}

The unique dependence of $I_J$ on $\phi$ and on the orientation of
${\bf M}$ can be used to build a Josephson current switch in which
$I_J$ is turned ``on" or ``off" via the rotation of ${\bf M}$. This
is shown in Fig.~\ref{parallel_d_alpha}(a) where we present $I_J$ as
a function of $\alpha$ for ${\bf d}_{L}
\parallel {\bf d}_{R}$ at $T=0$.
%
%
Note that the $\alpha$-dependence of $I_J$ changes significantly
with increasing $\phi$. In particular, $I_J$ becomes sharply peaked
around $\alpha=(2n+1)\pi/2$ with integer $n$ (i.e., for ${\bf M}
\perp {\bf d}_{L,R}$) as $\phi$ approaches $\pi$. As a result, small
variations in $\alpha$ lead to large changes in the magnitude of the
Josephson current, and can thus tune the junction between an
``on"-state ($I_J \not = 0$) and an ``off"-state ($I_J \approx 0$).
This behavior of $I_J$ is directly reflected in the
$\alpha$-dependence of the bound state energies, as shown in
Fig.~\ref{parallel_d_alpha}(b). While for $\phi=0$ the Andreev
states are $\alpha$-independent (see above), $E_{a,b}$ oscillate
sinusoidally with $\alpha$ for $\phi=\pi/4$, as shown in
Fig.~\ref{parallel_d_alpha}(b). As $\phi$ approaches $\pi$ (see,
e.g., $\phi=0.9 \pi$), the energies of the Andreev states alternate
in being close to zero and almost $\alpha$-independent for half of
the period and sinusoidal for the other half. Finally, for
$\phi=\pi$ (not shown) the bound state energies are given by
$E_a/\Delta_0= 2k_F\sqrt{D(1-D)}\cos{\alpha}$ and $E_b=0$. Note that
for $\phi \leq \pi$, the Andreev states exhibit no zero-energy
crossings when ${\bf M}$ is rotated .

\begin{figure}[h]
\includegraphics[width=8.5cm,angle=0]{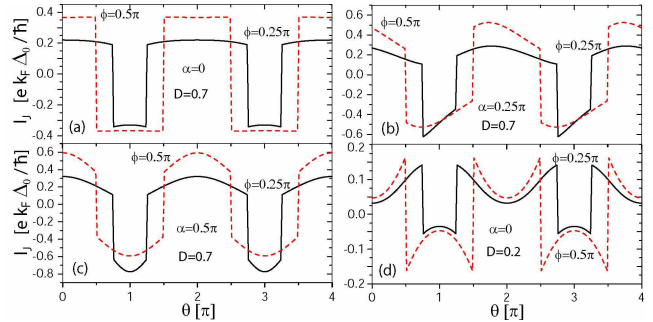}
\caption{$I_J$ as a function of $\theta$ for $T=0$, $D=0.7$, several
values of $\phi$, and (a) $\alpha=0$, (b) $\alpha=\pi/4$, and (c)
$\alpha=\pi/2$. (d) $I_J$ for $D=0.2$ and $\alpha=0$.}
\label{non_parallel_theta}
\end{figure}

Another type of Josephson switch can be created by rotating ${\bf
d}_R$ for fixed ${\bf M}$ and ${\bf d}_L$, as shown in
Fig.~\ref{non_parallel_theta} for $T=0$.
%
%
For $\alpha=0$, $\phi=\pi/2$ and $D=0.7$, $I_J$ exhibits an almost
perfect square wave form, is symmetric around $\theta=n \pi$, and is
nearly $\theta$-independent between the discontinuous jumps, as
shown in Fig.~\ref{non_parallel_theta}(a). For $\phi \not = \pi/2$,
this symmetry is broken, and the ranges of $\theta$ over which $I_J$
is positive or negative are unequal, but the discontinuous jumps in
$I_J$ persist. For $\alpha=\pi/4$, the Josephson current is skewed,
and in some regions varies nearly linearly with $\theta$
[Fig.~\ref{non_parallel_theta}(b)]. For $\alpha=\pi/2$ and
$\phi=\pi/2$, $I_J$ is again symmetric around $\theta=n \pi$
[Fig.~\ref{non_parallel_theta}(c)]. With decreasing $D$, the
Josephson current becomes $\theta$-dependent between discontinuous
jumps, even for $\alpha=0$ and $\phi=\pi/2$
[Fig.~\ref{non_parallel_theta}(d)]. Thus small changes in the
relative alignment of the ${\bf d}$-vectors can switch the junction
between two {\it current states} with opposite direction of $I_J$.

\begin{figure}[h]
\includegraphics[width=8.5cm,angle=0]{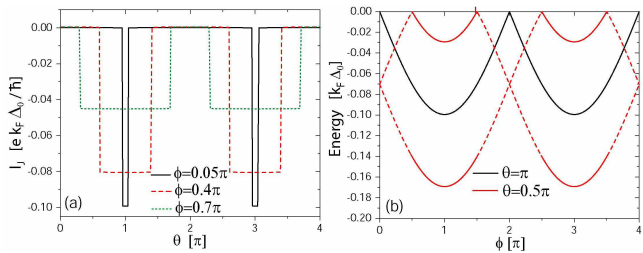}
\caption{Correlated rotation of ${\bf M}$ and ${\bf d}_R$ with
$\alpha=\theta/2$ and $D=0.01$ at $T=0$. (a) $I_J$ as a function of
$\theta$, and (b) $-E_{a,b}$ as a function of $\phi$.} \label{corr}
\end{figure}

A third type of Josephson switch can be formed via a correlated
rotation of ${\bf M}$ and ${\bf d}_R$. Specifically, we note that
$E_{a,b}$, Eq.(\ref{Eb}), depends on $\alpha$ only via
$\cos^2(\alpha-\theta/2)$. We therefore consider a simultaneous
rotation of ${\bf d}_R$ and ${\bf M}$  such that $\alpha=\theta/2$,
and present the resulting $I_J$ as a function of $\theta$ in
Fig.~\ref{corr}(a).
%
%
$I_J$ then exhibits square wave oscillations with discontinuous
jumps from $I_J \approx 0$ to a negative (positive) value for $0
\leq \phi \leq \pi$ ($\pi \leq \phi \leq 2\pi$). To understand these
sharp transitions, we plot the $\phi$-dependence of $-E_{a,b}$ in
Fig.~\ref{corr}(b) (only these branches contribute to $I_J$ at
$T=0$). For $\theta=\pi$ the Andreev states are degenerate, with
$\partial E_{a,b}/\partial \phi \leq 0 (\geq 0)$ for $0 \leq \phi
\leq \pi$  ($\pi \leq \phi \leq 2\pi$). In contrast, for $\theta=
\pi/2$ the degeneracy of the Andreev states is lifted with $\partial
E_a/\partial \phi \approx -\partial E_b/\partial \phi$ in some range
of $\phi$ (shown as dashed lines in Fig.~\ref{corr}(b)) such that
$I_J \approx 0$, but $I_J \not = 0$ in other regions of $\phi$
(shown as solid lines). Note that as $\phi$ increases from zero, the
range in $\theta$ in which $I_J \not = 0$ increases, while the
magnitude of $I_J$ decreases. This correlated rotation thus
represents another type of Josephson current switch in which $I_J$
can be turned ``on" or ``off".

Above, we computed $I_J$ assuming that the phase difference between
the superconductors, $\phi$, or the relative orientations between
${\bf M}$ and ${\bf d}_{L,R}$ (i.e., $\alpha$ and $\theta$) are
maintained for a sufficiently long time such that the system is in
thermodynamic equilibrium. The qualitative form of $I_J$ changes,
however, if ${\bf M}$ or ${\bf d}_{L,R}$ are rotated adiabatically
on a time scale which is shorter than the relaxation time required
for the occupation of the Andreev states to achieve their
equilibrium values. In this case, the occupation of the Andreev
states remains unchanged from an initial value during the adiabatic
rotation \cite{Bas99}.
%
%
\begin{figure}[h]
\includegraphics[width=8.5cm,angle=0]{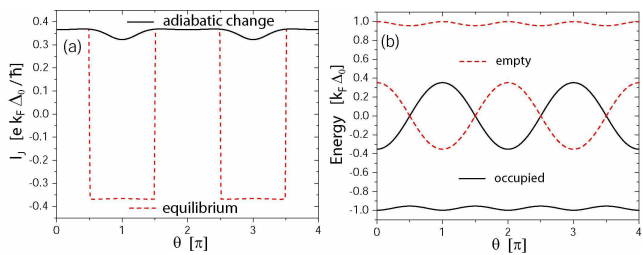}
\caption{(a) $I_J$ at $T=0$ for an adiabatic rotation of ${\bf M}$
and in equilibrium (see also Fig.~\ref{non_parallel_theta}(a)) as a
function of $\theta$. $\alpha=0$, $\phi=\pi/2$ and $D=0.7$ . (b)
$\pm E_{a,b}$ as a function of $\theta$. The black solid lines
represent the occupied states if $\theta$ is changed adiabatically.}
\label{NEQ}
\end{figure}
The Josephson current at $T=0$ for an adiabatic rotation of ${\bf
d}_{R}$ and in the equilibrium case are shown in Fig.~\ref{NEQ}(a).
To understand the {\it qualitative} form of $I_J$, we note that in
the equilibrium case, only the negative energy branches of the
Andreev states contribute to $I_J$. In contrast, if $\theta$ is
adiabatically changed (from the state with $\theta=0$ that is in
thermodynamic equilibrium), then the energy branches of the Andreev
states represented by the black solid lines in Fig.~\ref{NEQ}(b)
remain occupied, while those shown as red dashed lines remain
unoccupied. As a result, $I_J$ in the adiabatic case changes
continuously with $\theta$ and does not exhibit the discontinuous
jumps shown by the equilibrium $I_J$. Note that $I_J$ is
$2\pi$-periodic in both cases. A special case arises for
$\alpha=\pi/2$, when the periodicity of $I_J$ is changed to $4\pi$
(not shown). An adiabatic change of $\phi$ also leads to a
$4\pi$-periodicity of $I_J$, in analogy to the fractional
$ac$-Josephson effect discussed in Ref.~\cite{Kwon04}.

Finally, scattering off the ferromagnetic barrier in general leads
to a suppression of the superconducting order parameter near the
barrier, which was not accounted for in the above approach. However,
in nodeless superconductors, such as the one discussed above, the
order parameter recovers its bulk value on a length scale set by
$1/k_F$ \cite{suppression}, which is assumed to be much shorter than
the decay length of the Andreev bound state \cite{Kwon04,Zag98}. As
a result, the spatial variation of the order parameter leads to only
weak quantitative and no qualitative changes in the induced
fermionic bound states \cite{suppression}. We thus expect that the
results presented above are unaffected by the inclusion of the order
parameter suppression.

In summary, we predict a new type of Josephson effect in TSFTS
junctions, in which $I_J$ exhibits a rich dependence on the relative
orientation of ${\bf M}$ and ${\bf d}_{L,R}$. This dependence can be
used to build several types of {\it Josephson current switches} in
which small changes of $\alpha$ or $\theta$ can tune the junction
between two {\it current states}. We predict an unconventional
temperature dependence of $I_J$ such that for certain orientations
of ${\bf M}$ and ${\bf d}_{L,R}$, $I_J$ changes sign with increasing
temperature.

We are grateful to F. Nogueira and A. Sudb{\o} for helpful
discussions. D.K.M. acknowledges financial support from the
Alexander von Humboldt foundation.


\begin{thebibliography}{99}
\bibitem{com2} The Josephson current carried by states outside the
superconducting gap is in general negligible \cite{Rie98,Zag98}.
\end{thebibliography}
\end{document}